\documentclass[aps,pra,showpacs,groupedaddress,twocolumn,showkeys]{revtex4-2}
\usepackage{amsmath}
\usepackage{graphicx}
\usepackage{diagbox}
\usepackage{xcolor}
\usepackage[colorlinks=true, letterpaper=true, pdfstartview=FitV, linkcolor=blue, citecolor=blue, urlcolor=blue]{hyperref}

\begin{document}

\title{Flat-top solitons and anomalous interactions in media with even-order dispersions and competing nonlinearities}

\author{Xueqing He$^{1,2,3}$}
\author{Shijie Hao$^{1,2}$}
\author{Lijing Xing$^{1,2}$}
\author{Dumitru Mihalache$^{4}$}
\author{Boris A. Malomed$^{5,6}$}
\author{Pengfei Li$^{1,2,3}$}
\email{lpf281888@gmail.com}

\affiliation{$^{1}$Department of Physics, Taiyuan Normal University, Jinzhong, Shanxi 030619, China}
\affiliation{$^{2}$Shanxi Key Laboratory for Intelligent Optimization Computing and Blockchain Technology, Taiyuan Normal University, Jinzhong, Shanxi 030619, China}
\affiliation{$^{3}$Institute of Computational and Applied Physics, Taiyuan Normal University, Jinzhong, Shanxi 030619, China}
\affiliation{$^{4}$Horia Hulubei National Institute of Physics and Nuclear Engineering, Magurele, Bucharest RO-077125, Romania}
\affiliation{$^{5}$Department of Physical Electronics, School of Electrical Engineering, Faculty of Engineering, and Center for Light-Matter Interaction, Tel Aviv University, Tel Aviv 69978, Israel}
\affiliation{$^{6}$Instituto de Alta Investigaci\'{o}n, Universidad de Tarapac\'{a}, Casilla 7D, Arica, Chile}

\begin{abstract}
Flat-top (FT) solitons are optical pulses that arise from the balance of dispersion and self-phase modulation in media with the competing cubic-quintic nonlinearity. Previously, FT solitons were studied only in the case of the second-order dispersion ($m=2$).
Following the recent observation of pure-quartic solitons (corresponding to $m=4$), we here construct families of FT solitons in the setting with pure-high-even-order dispersion (PHEOD), including $m=4,6,8$, and $10$, and address interactions between them. The PHEOD solitons are completely stable, and, unlike the conventional solitons, they feature oscillatory tails. Interactions between the PHEOD solitons are anomalous, featuring repulsion and attraction between in- and out-of-phase solitons, respectively. These results expand the variety of optical solitons maintained by diverse dispersive nonlinear media.
\end{abstract}

\maketitle

\section{Introduction}

Much interest to flat-top (FT) solitons was drawn by the experimental
realization of quantum droplets \cite{Pfau_2021,Luo2020} in dipolar \cite%
{Kadau2016,Schmitt2016,PhysRevX.6.041039} and binary \cite%
{PhysRevLett.120.135301,doi:10.1126/science.aao5686,PhysRevLett.120.235301,PhysRevResearch.1.033155}
Bose-Einstein condensates (BECs) in ultracold atomic gases. FT solitons are
produced by nonlinear Schr\"{o}dinger equations (NLSEs) with competing
nonlinearities. In BEC mixtures, competing nonlinearities originate from the
interplay of the mean-field inter-atomic interactions and Lee-Huang-Yang
correction induced by quantum fluctuations \cite%
{PhysRev.105.1119,PhysRev.106.1135}, which leads to the creation of FT modes
in the form of quantum droplets \cite{Petrov1,Petrov2}. In nonlinear optics,
competing nonlinearities usually combine the self-focusing cubic and
defocusing quintic terms, which provides the stabilization of two- and
three-dimensional solitons against the collapse driven by the self-focusing
\cite{Bulgaria,Quiroga,Pego,PhysRevE.61.3107,PhysRevE.65.066604,PhysRevLett.88.073902,PhysRevE.69.056601,PhysRevLett.103.023903}
.

Recently, a new class of soliton pulses, known as pure-quartic solitons
(PQSs), has been identified, arising from the balance between the
fourth-order group-velocity dispersion (GVD) and self-phase modulation
(SPM). PQSs have been predicted and experimentally observed in
photonic-crystal waveguides by means of precise GVD engineering \cite%
{blanco2016pure}, using a mode-locked laser incorporating an intra-cavity
spectral pulse shaper \cite{runge2020pure,lpor.202501851}. The conventional
solitons, supported by the second-order GVD, and PQSs are just the two
lowest-order members of an infinite hierarchy of soliton pulses arising from
the interplay of the Kerr nonlinearity and negative pure-high-even-order
(PHEOD) dispersion. Experimental studies indicate that more general PHEOD
solitons arise from the balance between the SPM and a single negative PHEOD
dispersion term of orders $m=6$, $8$, and $10$ \cite{PhysRevResearch.3.013166}. In subsequent research, PHEOD solitons have been demonstrated to form bound states \cite{han2024pure,LI2026117742}.

Unlike the conventional solitons, PQSs and PHEOD solitons feature
oscillations in their exponentially decaying tails \cite{Tam:19}.
Importantly, the PQSs obey the scaling relation with the energy proportional
to the third power of the inverse pulse duration, compared to the
conventional solitons, whose energy scales as the inverse of the pulse
duration, $\tau _{0}^{-1}$. Generally, the energy of PHEOD solitons scales
as $\tau _{0}^{-(m-1)}$, implying that they obey a favorable energy scaling,
making it possible to attain higher energies than the conventional GVD
solitons. Another difference is that the PQSs and PHEOD solitons do not obey
the Galilean invariance, making the creation of moving solitons a nontrivial
issue \cite{PhysRevA.104.043526}.

Recent advances in the work with PQSs and PHEOD solitons suggest a
possibility for the development of a new branch of nonlinear optics and may
lead to novel applications \cite{10.1063/5.0059525,luo2023research}.
Recently, several species of such solitons have been numerically
investigated in the framework of the NLSE, including dissipative \cite%
{Wu:25,Silvestri:25}, Raman \cite{Liu:21,Wang:22,Zhu:25}, dark \cite%
{Alexander:22,PhysRevA.111.013510}, and spatiotemporal pure-quartic \cite%
{PhysRevA.109.033516} solitons, as well as pure-quartic domain walls \cite%
{Li:25} and bound states of PQSs (``molecules") \cite%
{Deng:25,PhysRevA.111.063503}. However, PQSs and PHEOD solitons of the FT
type have not yet been investigated. In this work we focus on FT solitons
produced by the NLSE with the PHEOD terms and competing cubic-quintic (CQ)
nonlinearity. Our objective is to reveal properties of the PHEOD solitons of
the FT type and their interactions. By means of numerically methods, we
produce families of FT solitons supported by GVD of orders $m=4$, $6$, $8$,
and $10$. Our study shows anomalous interactions of such solitons. These
findings help to further understand the physical purport of the PHEOD
solitons and extend the variety of optical solitons in dispersive nonlinear
media.

\section{Model and Methods}

The starting point is the modified one-dimensional NLSE\ which governs the
propagation of optical pulses in the medium with PHEOD and the CQ
nonlinearity:
\begin{equation}
i\frac{\partial \Psi }{\partial z}-\frac{\partial ^{m}\Psi }{\partial \tau
^{m}}+|\Psi |^{2}\Psi -|\Psi |^{4}\Psi =0,  \label{NLSE1}
\end{equation}%
where $\Psi \left( z,\tau \right) $ is the field envelope, $z$ is the
propagation distance, and $\tau $ is the retarded time in the reference
frame moving with the group velocity of the carrier wave. We consider the
anomalous sign of PHEOD of even orders $m=4,6,8,10$.

We look for stationary PHEOD solutions to Eq. (\ref{NLSE1}) with a real
temporal profiles $\psi $ and propagation constant $\beta $ as
\begin{equation}
{\Psi \left( z,\tau \right) }=e^{i\beta z}{\psi (\tau )}.  \label{Psi}
\end{equation}%
The substitution of this in Eq. (\ref{NLSE1}) leads to the equation for $%
\psi $:
\begin{equation}
-\frac{d^{m}\psi }{d\tau ^{m}}+|\psi |^{2}\psi -|\psi |^{4}\psi -\beta \psi
=0.  \label{system_psi1}
\end{equation}%
We solved Eq. (\ref{system_psi1}) numerically, using the Newton-conjugate
gradient method \cite{yang2010nonlinear}. With an appropriate initial guess,
the method converges to numerical solutions through successive iterations.

\section{Numerical Results}

\begin{figure}[th]
	\centering
	\includegraphics[width=1\linewidth]{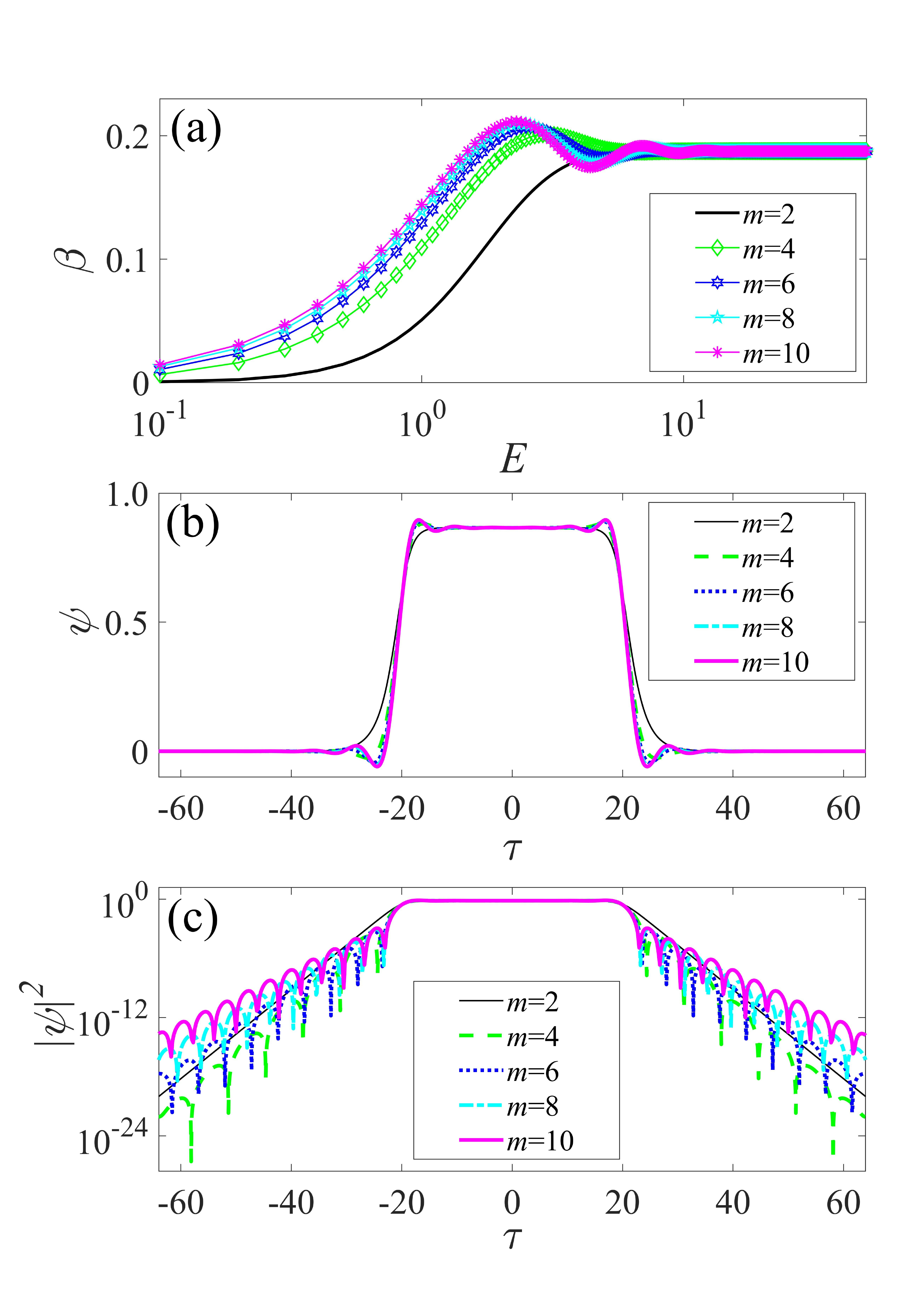}
	\caption{Families of numerically found PHEOD solitons of the FT type. (a)
		The propagation constant vs. the soliton's energy (shown on the logarithmic
		scale) for GVD$\ $orders $m=4,6,8,10$, compared to the $\protect\beta (E)$
		curve for conventional FT solitons, for $m=2$. (b) Profiles of the PHEOD and
		conventional FT solitons with a fixed energy, $E=30$. (c) The same as in
		(b), but on the logarithmic scale, to display oscillatory tails of the PHEOD
		FT solitons.}
	\label{fig1}
\end{figure}

Families of PHEOD solitons are characterized by their energy (alias the
integral norm),
\begin{equation}
E(\beta )=\int_{-\infty }^{+\infty }|\psi |^{2}d\tau,  \label{Power}
\end{equation}%
which is a dynamical invariant of Eq. (\ref{NLSE1}). The numerically
obtained dependence of their propagation constant $\beta $ on energy $E$ is
plotted, on the logarithmic scale, in Fig. \ref{fig1}(a). To compare FT
solitons produced by Eq. (\ref{system_psi1}) with the conventional FT
solitons, known from the solution of the second-order NLSE with the CQ
nonlinearity, the curve $\beta (E)$ for $m=2$ is included too, showing that
it is essentially different from its counterparts for $m\geq 4$. In the
latter case, the propagation constant does not increase monotonously with
the increase of the soliton's energy, but instead exhibits oscillations
before approaching the commonly known limit value $\beta _{\mathrm{lim}}=3/16
$, determined by the CQ nonlinearity. The limit value is the universal 
one for the cubic-quintic nonlinearity, which does not depend on the dispersion,
being actually determined by the uniform (continuous-wave) states. In the rigorous 
form, the existence of the limit value was proven with the help of the Gagliardo-Nirenberg and H\"{o}lder 
inequalities, together with the Pohozhaev identities \cite{PhysRevE.78.027601}.

\begin{figure}[th]
	\centering
	\includegraphics[width=1\linewidth]{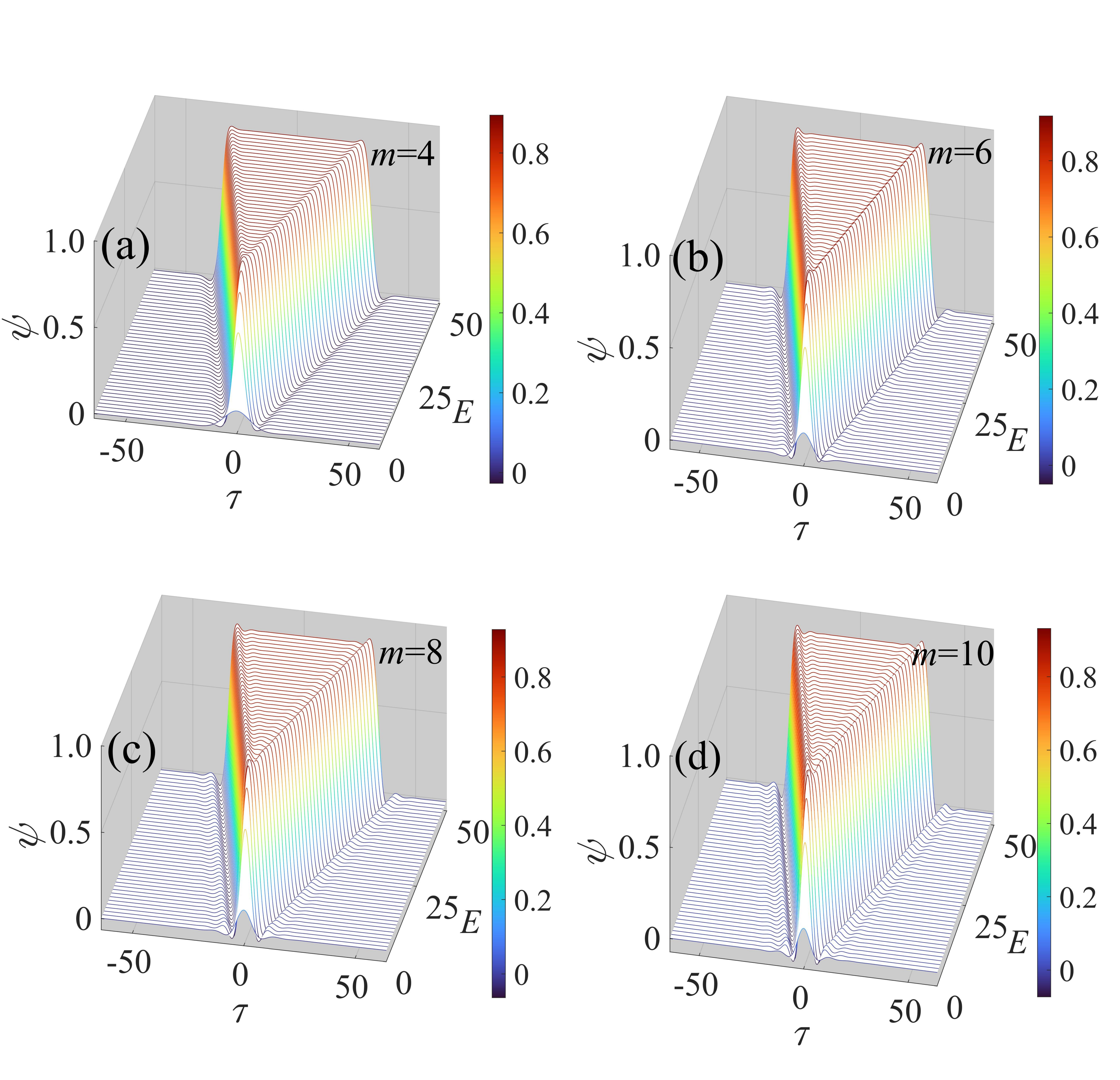}
	\caption{The variation of profiles of the PHEOD solitons, produced by the
		numerical solution of Eq. (\protect\ref{system_psi1}) with $m=4$, $6$, $8$,
		and $10$, for the soliton's energy taking values $0.1\leq E\leq 50$ with
		interval $\Delta E=0.1$.}
	\label{fig2}
\end{figure}

Typical examples of the temporal profiles of the PHEOD FT solitons, for a
fixed energy, $E=30$, are plotted in Figs. \ref{fig1}(b) and \ref{fig1}(c).
As seen in Fig. \ref{fig1}(c), the profiles exhibit significant oscillating
tails, whose decay rate and oscillation period of the tails are determined
by the linearization of Eq. (\ref{system_psi1}). Further, profiles of the
PHEOD FT solitons are displayed in Fig. \ref{fig2}, for the energy $E$
ranging from $0.1$ to $50$, which indeed demonstrates the trend to the
formation of FT shapes of the solitons in the limit of large energies.

It is worth noting that the flat-top solitons considered in the present paper 
differ from platicon modes produced the Lugiato-Lefever equation \cite{Lobanov:15}. 
Specifically, the flat-top solitons are bright states formed by the competition between the 
cubic self-focusing and quintic defocusing, whereas platicons are dissipative dark solitons. 
In optical microcavities, platicons are generated via the double balance mechanism: between 
the normal GVD and Kerr self-focusing, and also between the gain and loss.

\begin{figure}[th]
	\centering
	\includegraphics[width=1\linewidth]{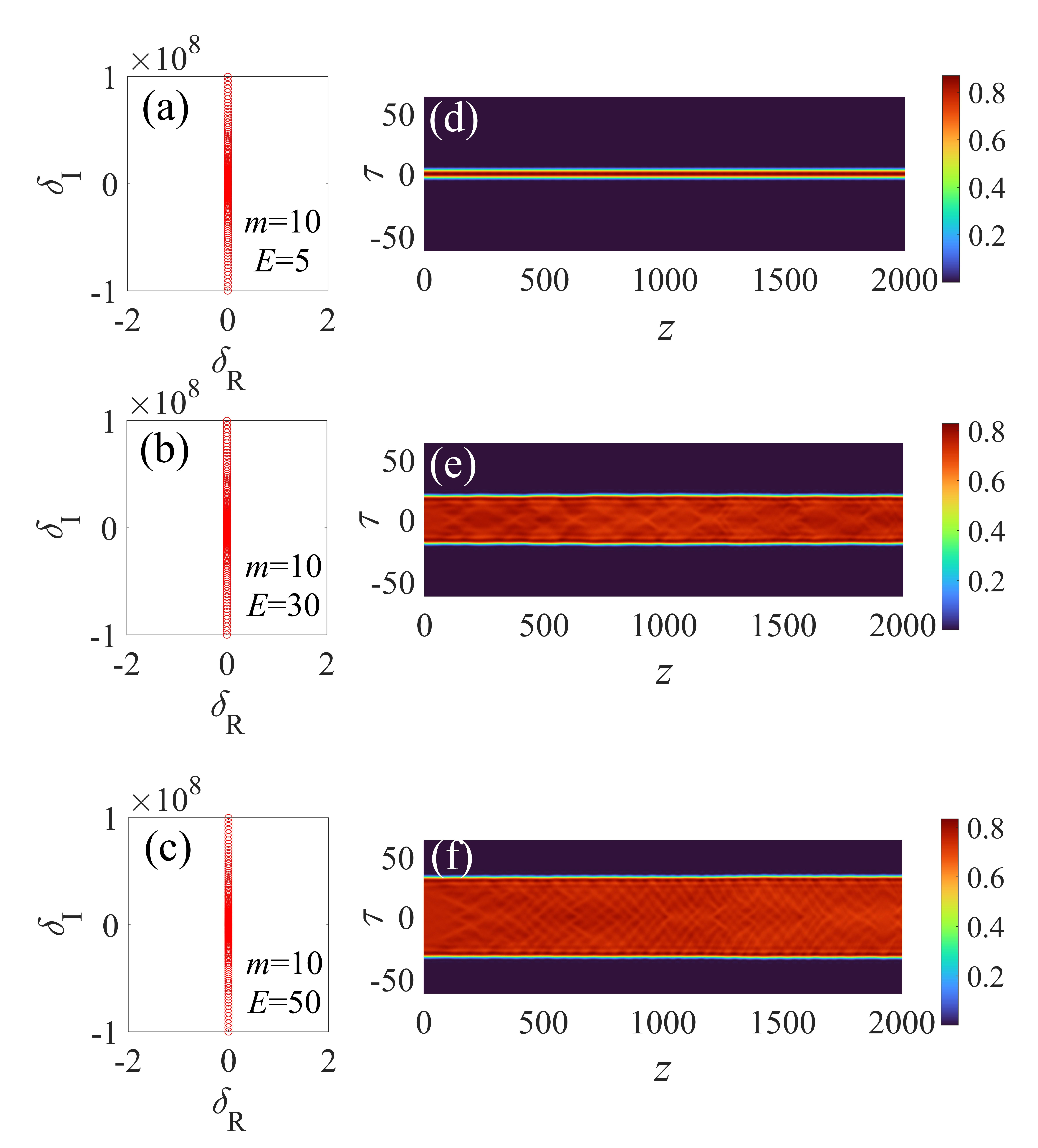}
	\caption{Linear-stability spectra and the perturbed evolution of the
		solitons for the GVD order $m=10$. The linear-stability spectra are plotted
		in the left panels with different energies: (a) $E=5$, (b) $E=30$, and (c) $%
		E=50$, while the right panels (d-f) show the correponding stable evolution
		with initial random perturbations at a $5\%$ amplitude level.}
	\label{fig3}
\end{figure}

To test the stability of the solitons, we computed eigenvalues and
eigenmodes of small perturbations added to the solitons, using the Fourier
collocation method \cite{yang2010nonlinear}, and simulating the perturbed
evolution. To this end, perturbed solutions were introduced as
\begin{equation}
\Psi =e^{i\beta z}\left[ \psi \left( \tau \right) +u\left( \tau \right)
e^{\delta z}+v^{\ast }\left( \tau \right) e^{\delta ^{\ast }z}\right] ,
\label{Perturbation1}
\end{equation}%
where $\psi $ represents the unperturbed soliton, defined as per Eq. (\ref%
{Psi}), while $u$ and $v$ are small perturbations with the respective
complex eigenvalue $\delta $, and $\ast $ standing for the complex
conjugate. The substitution of ansatz (\ref{Perturbation1}) in Eq. (\ref%
{NLSE1}) and linearization leads to the system of coupled equations:

\begin{equation}
\delta u=+iLu+i\psi^{2}v-2i\psi^{2}|\psi|^{2}v,  \label{Linearization_a}
\end{equation}

\begin{equation}
\delta v=-iLv-i\psi^{2}u+2i\psi^{2}|\psi|^{2}u,
\label{Linearization_b}
\end{equation}

\begin{equation}
L\equiv -\beta -\frac{d^{m}}{d\tau ^{m}}+2|\psi |^{2}-3|\psi |^{4},
\label{Linearization_L}
\end{equation}%
which were solved by means of the Fourier collocation method. The solitons
are unstable if there are eigenvalues with $\mathrm{Re}(\delta )>0$.

The results demonstrate that the PHEOD solitons of the FT type are stable,
see typical examples of the linear-stability spectra in Figs. \ref{fig3}%
(a-c). Direct stability tests were performed by simulations of Eqs. (\ref%
{NLSE1}), using the Runge-Kutta method. Typical examples of the stable
evolution of the PHEOD FT solitons are displayed in Figs. \ref{fig3}(d-f).
The results confirm that they remain stable, at least, up to $z=2000$, under
the action of random initial perturbations with a relatively large amplitude
of $5\%$.

\begin{figure}[th]
	\centering
	\includegraphics[width=1\linewidth]{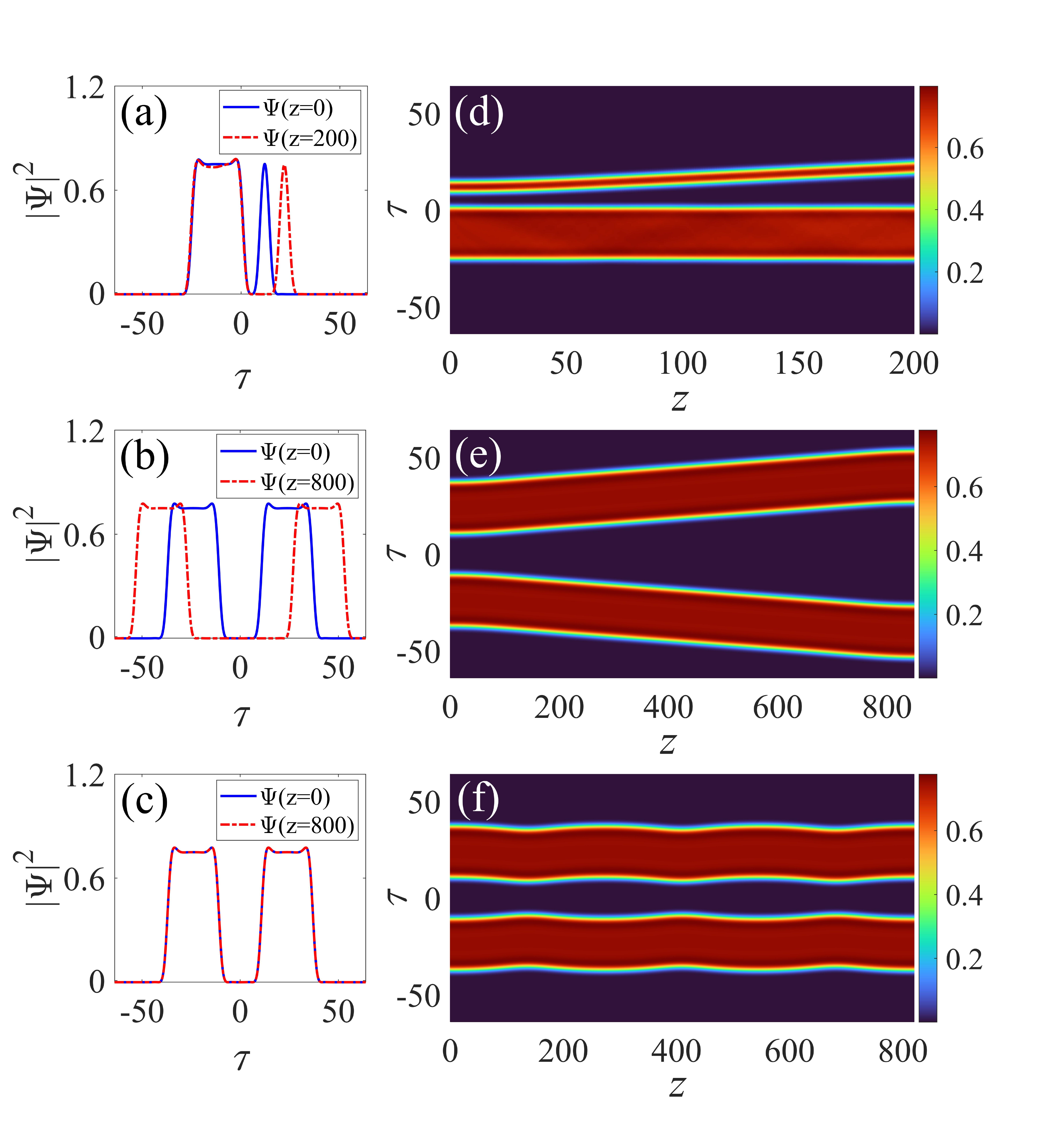}
	\caption{Interactions of the pure-decic flat-top solitons with different
		initial relative phase. Intensity input (solid lines), output (dashed lines)
		profiles and evolutions of the interacting solitons for the following cases:
		(a) and (d) interactions of ${\Psi_1(\mathit{E}=20)}$ and ${\Psi_2(\mathit{E}%
			=4)}$ with the initial relative phase $\Delta\protect\phi=0$, (b) and (e)
		interactions of ${\Psi_1(\mathit{E}=20)}$ and ${\Psi_2(\mathit{E}=20)}$ with
		the initial relative phase $\Delta\protect\phi=0$, (c) and (f) interactions
		of ${\Psi_1(\mathit{E}=20)}$ and ${\Psi_2(\mathit{E}=20)}$ with the initial
		relative phase $\Delta\protect\phi=\protect\pi$.}
	\label{fig4}
\end{figure}

Finally, we address interactions of the PHEOD FT solitons, using the
following ansatz to represent two weakly overlapping solitons:
\begin{equation}
{\Psi \left( z,\tau \right) }={\Psi _{1}(\tau _{+})}+{\Psi _{2}(\tau _{-})}%
e^{i\Delta \phi }.  \label{Psi_in}
\end{equation}%
Here ${\tau _{\pm }\equiv \tau \pm \Delta \tau }$, with $2\Delta \tau $ and $%
\Delta \phi $ standing, respectively, for the separation and phase shift
between the solitons.

To illustrate the interaction of PHEOD solitons, the evolution of two
in-phase ones with initial half-separation $\Delta \tau =12$ and $24$ is
displayed in Figs. \ref{fig4}(a,d) and (b,e), respectively, which exhibit
repulsion between the in-phase PHEOD solitons (in the former case, one
soliton is broad (FT) and one narrow, while in the latter case both are
broad). On the other hand, out-of-phase FT PHEOD solitons attract each other,
as seen in Figs. \ref{fig4}(c,f). The character of these interactions is
anomalous, as the sign of the interaction is exactly opposite to that
featured by the conventional Kerr solitons \cite%
{doi:10.1126/science.286.5444.1518}.

\section{Conclusion}

In conclusion, we have reported the existence of the new type of completely
stable\ FT (flat-top) solitons in optical media combining PHEOD
(pure-high-even-order dispersion), of orders $m=4$, $6$, $8$, $10$, and the
competing CQ (cubic-quintic) nonlinearity. Unlike the traditional solitons
corresponding to $m=2$, the FT PHEOD solitons feature oscillatory tails, and
the respective soliton families are characterized by the non-monotonous
dependence of the propagation constant on the total energy, demonstrating
the growth with oscillations. Interactions between the PHEOD solitons are
anomalous, featuring repulsion between in-phase solitons and attraction
between out-of-phase ones. Thus, our findings expand the understanding of
temporal PHEOD solitons in optics. As an extension of the present work, it
is interesting to study in detail moving states of the novel solitons, as
well as their bound states and interaction between the bound states.

\begin{acknowledgments}
This research was supported by the National Natural Science Foundation of China (11805141); Fundamental Research Program of Shanxi Province (202303021211185); Israel Science Foundation (grant No. 1695/22).
\end{acknowledgments}

\end{document}